\documentclass[a4paper]{mem}
\usepackage{natbib}
\usepackage{graphicx}
\usepackage[a4paper]{hyperref}
\idline{00}{00} 
\def\xmm{{\sl XMM}} 
\def\cha{{\sl Chandra}}
\newcommand{\gapr}{\raisebox{-.6ex}{\mbox{
$\stackrel{>}{\mbox{\scriptsize$\sim$}}\:$}}}
\newcommand{\lapr}{\raisebox{-.6ex}{\mbox{
$\stackrel{<}{\mbox{\scriptsize$\sim$}}\:$}}}
\begin{document}
\title{XMM observations of three middle-aged pulsars
}

   \author{V.~E. Zavlin \inst{1} 
          \and
          G.~G. Pavlov \inst{2}
}


   \institute{Max-Planck Institut f\"ur extraterrestrische Physik, 
85748 Garching, Germany
         \and
Dept.\ of Astronomy and Astrophysics, Pennsylvania State
University, 525 Davey Lab, University Park, PA 16802, USA
             }
   \abstract{
X-ray observations of middle-aged pulsars
allow one to study nonthermal radiation from pulsar magnetospheres
and thermal radiation from neutron star (NS) surfaces.
In particular, from the analysis of thermal radiation one can infer
the surface temperatures and radii of NSs, which 
is important for investigating evolution of these objects and 
constraining the equation of state of the superdense matter
in the NS interiors.  Here we present results of \xmm\ observations 
of three middle-aged pulsars,
J0538+2817, B0656+14 and J0633+1746 (Geminga),
and briefly discuss mechanisms of their X-ray emission.

   }
   \authorrunning{V.E Zavlin and G.G. Pavlov}
   \titlerunning{\xmm\ observation of middle-aged pulsars}
   \maketitle

\section{Introduction}
A large number of isolated neutron stars (NSs) of various types,
including active radio pulsars, 
have been observed with the currently
operating large X-ray observatories, \cha\ and \xmm.
Recent reviews on \cha\ observations of NSs can be found in 
Pavlov et al.\ (2002; [PZS02]) and Weisskopf (2002),
while first \xmm\ results are described by
Becker \& Aschenbach (2002).
Here we discuss \xmm\ observations of three pulsars, J0538+2817, B0656+14 
(J0538 and B0656, hereafter) and J0633+1746 (Geminga), with 
ages of 30--300 kyr and spin-down energy loss rates of 
(3--$5)\times 10^{34}$ erg s$^{-1}$ (see Table 1).

X-ray emission from an active pulsar
generally consists of two components, thermal and
nonthermal. In very young, bright pulsars 
($\tau < 10$ kyr) the thermal component is buried under the powerful 
nonthermal emission,
while X-ray emission from old pulsars ($\tau\gapr 10^3$ kyr)
is usually very faint.
The intermediate group of middle-aged pulsars 
is particularly interesting for X-ray studies because 
they are expected to be bright and exhibit both thermal and
nonthermal components.
Below we present results of spectral and timing analysis of 
three middle-aged pulsars observed with the
EPIC-pn instrument on board \xmm.

\begin{table*}
\begin{center}
\begin{tabular}{c|cccccccc}
\hline
Pulsar & P  & $\tau$ & $\dot{E}$  & $B$  & $d$ & $F_X$ & $L_X$ & $L_{\rm bol}$\\
      & ms &  kyr & $10^{34}$\,erg\,s$^{-1}$  &  $10^{12}$\,G & kpc & 
$10^{-12}$\,cgs & $10^{32}$\,erg\,s$^{-1}$ &$10^{32}$\,erg\,s$^{-1}$ \\
\hline
J0538  & 143  & $\,\,30$  & 5  & 1  & 1.2 & 0.6 & 6.5 & 11.2 \\
B0656    & 385  & 110  & 4  & 5  & 0.3 & 7.1 & 1.3 & $\,\,\,1.9$ \\      
Geminga     & 237  & 340  & 3  & 2  & 0.2 & 1.8 & 0.2 & $\,\,\,0.5$ \\
\hline\hline
\end{tabular}
\caption{Pulsar parameters: 
period $P$, age $\tau$ ($P/2\dot{P}$
for B0656 and Geminga, and true age for J0538 --- Kramer et al.\ 2003),
rotational energy loss $\dot{E}$, surface magnetic field $B$,
and distance $d$. Also, the energy flux 
$F_X$ in 0.2--10 keV (as detected with EPIC-pn),
the corresponding luminosity $L_X$ (corrected for interstellar absorption)
and the bolometric thermal luminosity $L_{\rm bol}$ are given.
}
\end{center}
\end{table*}

\section{PSR J0538+2817}
This radio pulsar was observed with \xmm\ in March 2002 
(11.3\,ks effective exposure for EPIC-pn, in small-window mode).
The detected spectrum of J0538 can be 
fitted with a single blackbody model of
$T^\infty_{\rm bb}\simeq 2.12$ MK,
without invoking a nonthermal (power-law) component 
(McGowen et al.\ 2003).
The blackbody model gives a small X-ray emitting area,
with a radius $R^\infty_{\rm bb}\simeq 1.7$ km\footnote{
All radii and luminosities
are given for the distances listed in Table 1.}, 
much smaller than the expected NS radius $\sim$10 km.
Such an area 
might be interpreted as a hot spot on the pulsar's surface. 
An alternative interpretation involves NS atmosphere models
(see Zavlin \& Pavlov 2002 [ZP02] for a review). 
A hydrogen atmosphere model with
$B=1\times 10^{12}$ G fits the data even better (Fig.\ 1) 
and yields a lower (unredshifted) surface temperature 
$T_{\rm eff}\simeq 1.12$ MK and a radius $R\simeq 10.5$ km, 
consistent with the expected NS radius.
An upper limit on the nonthermal X-ray 
luminosity in 0.1--10 keV is
$L^{\rm nonth}_X < 1\times 10^{31}$ erg s$^{-1}
=2\times 10^{-4}\dot{E}$.
Similar results were obtained 
by Romani \& Ng (2003) from 
a \cha\ observation,
albeit with somewhat different model parameters 
(the discrepancy is probably due to a pile-up effect in the \cha\ data).
They also found
an indication of a pulsar-wind nebula around J0538,
with a size of $\sim 18''$, that cannot be resolved
in the \xmm\ data. However, the estimated luminosity of the nebula,
$\sim 6\times 10^{-5}\dot{E}$, is too low to seriously affect the 
analysis of the EPIC-pn data.

   \begin{figure}
   \centering
   \includegraphics[width=9cm]{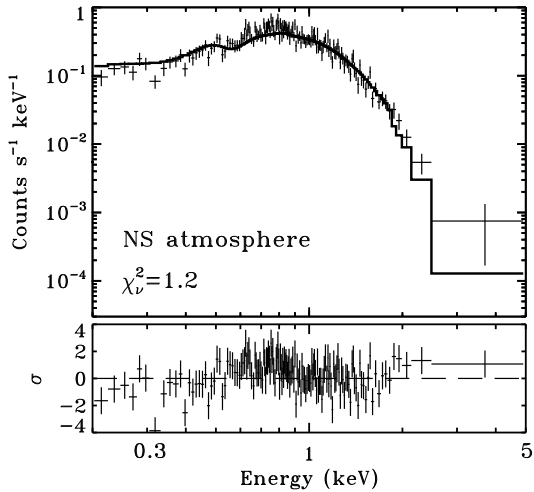}
      \caption{EPIC-pn count rate spectrum of J0538 fitted with a
 magnetized hydrogen atmosphere model.
              }
         \label{sp0538}
   \end{figure}

   \begin{figure}
   \centering
   \includegraphics[width=9cm]{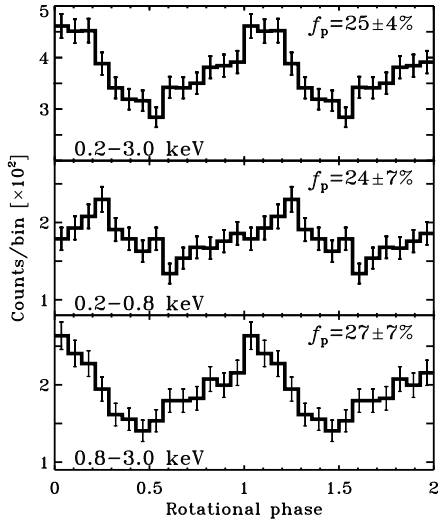}
      \caption{X-ray light curves of J0538.
              }
         \label{lc0538}
   \end{figure}

The EPIC-pn data also show pulsations of the 
X-ray flux, with one broad, asymmetric pulse per period 
(pulsed fraction $f_p\simeq 25\%$; McGowen et al.\ 2003). 
The phases of pulse maxima at energies below and above 0.8 keV 
differ by $\sim 75^\circ$ (Fig.\ 2). Such pulsations 
show that the thermal emission is intrinsically anisotropic and
indicate  strong nonuniformity of
the surface temperature and magnetic field.

\section{PSR B0656+14}

   \begin{figure}
   \centering
   \includegraphics[width=10cm]{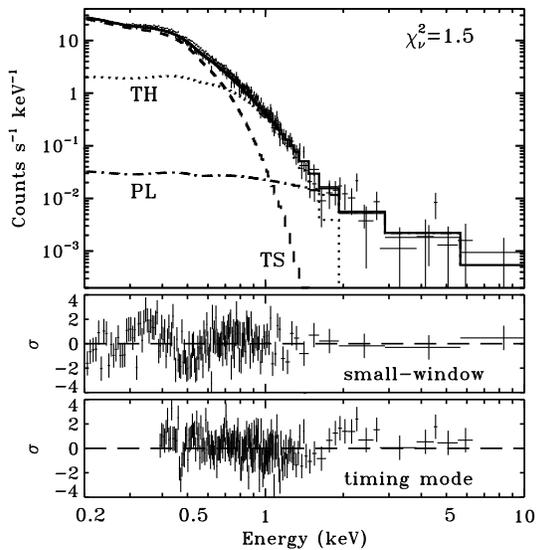}
      \caption{EPIC-pn count rate spectrum of B0656 taken in two observational
modes and fitted with a three-component model.
              }
         \label{sp0656}
   \end{figure}

   \begin{figure}
   \centering
   \includegraphics[width=9cm]{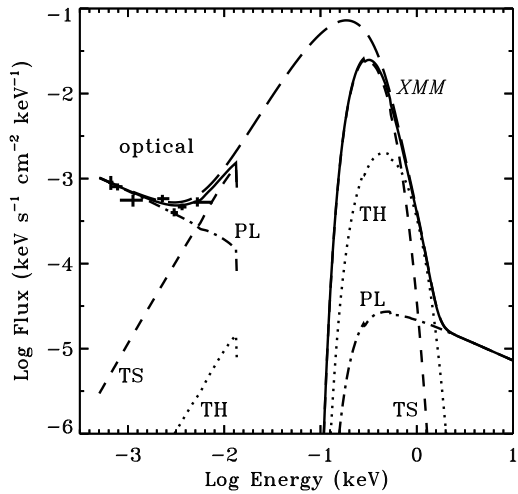}
      \caption{Multiwavelength spectrum of B0656, with 
extrapolations of the total
X-ray spectrum (long dashes) and its three components 
into the optical domain.
              }
         \label{msp0656}
   \end{figure}

   \begin{figure}
   \centering
   \includegraphics[width=13cm]{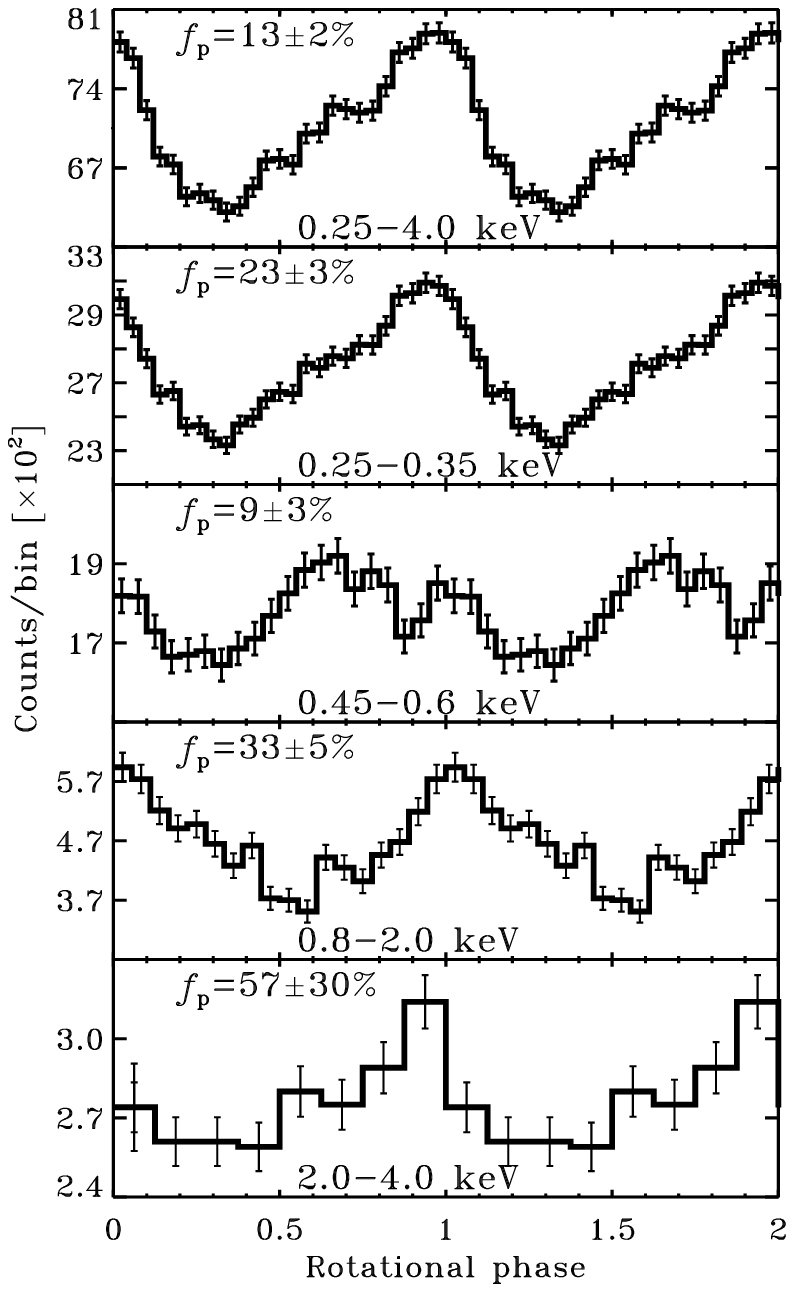}
      \caption{X-ray light curves of B0656. 
              }
         \label{lc0656}
   \end{figure}

B0656 was observed with \xmm\ in October 2001,
with 6.0 ks and 25.0 ks EPIC-pn exposures in small-window and
timing modes, respectively.
The results of the spectral analysis of the EPIC-pn data are 
quite consistent with those derived from \cha\ observations
(PZS02). The X-ray spectrum shows at least two distinct
components, thermal and nonthermal.
The best fit is provided by a model consisting of
soft (TS) and hard (TH) blackbody 
components, and a power law (PL; Fig.\ 3). The model parameters are 
$T^\infty_{\rm bb,s}\simeq 0.82$ MK, 
$R^\infty_{\rm bb,s}\simeq 7.3$ km (for TS), 
$T^\infty_{\rm bb,h}\simeq 1.72$ MK, 
$R^\infty_{\rm bb,h}\simeq 0.5$ km (for TH), 
and the photon index $\gamma\simeq 1.5$ (for PL).
This model fits well the multiwavelength spectrum of B0656,
from IR through X-ray energies (Fig.\ 4; see also PZS02
for details). However, even the 
larger radius, $R=[1+z]^{-1}R^\infty_{\rm bb,s}\approx 5$--6 km 
($z$ is the gravitational redshift), is too small
to be regarded as a NS radius.  Using NS hydrogen atmosphere models 
for the thermal component yields distances $d\simeq 0.1$ kpc 
(at $R=10$ km), a factor of three smaller than actually measured.
The nonthermal component contributes only 1\% of the
total luminosity in 0.1--10 keV, 
$L_X^{\rm nonth}\simeq 2\times 10^{30}\,{\rm erg}\,{\rm s}^{-1}=
5\times 10^{-5}\dot{E}$.

The timing analysis of the EPIC-pn data confirms 
that both the pulse
shape and pulsed fraction strongly depend on photon energy (Fig.\ 5).
The X-ray pulse profile of B0656 apparently consists of two components
(centered at $\phi\simeq 0.6$ and 1.0), whose relative contributions
depend on energy.  The nonthermal emission at
$E>2$ keV shows one narrow peak per period, with $f_p\gapr 30$\%. 
Unfortunately, strong background contamination 
(especially in timing mode) prevents 
a more accurate analysis of pulsations  
at these energies.

\section{Geminga}

   \begin{figure}
   \centering
   \includegraphics[width=9cm]{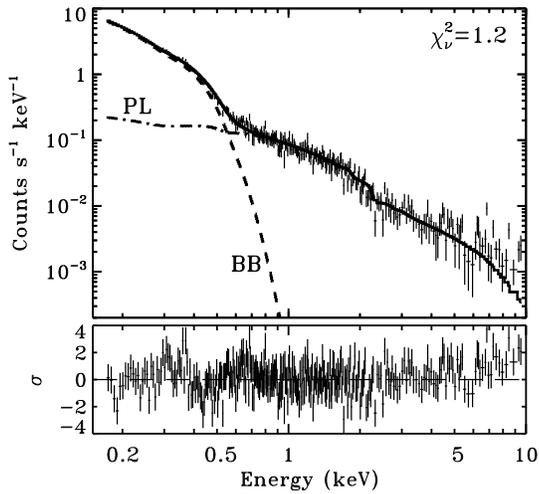}
      \caption{EPIC-pn count rate spectrum of Geminga
fitted with a two-component, blackbody (BB) plus
power-law (PL), model. 
}
         \label{spGem}
   \end{figure}

   \begin{figure}
   \centering
   \includegraphics[width=9cm]{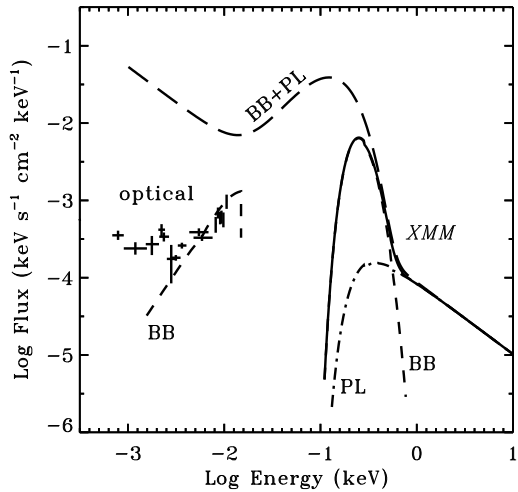}
      \caption{Multiwavelength spectrum of Geminga, with
extrapolations of the total
X-ray spectrum (long dashes) and its thermal component into the optical 
domain.
              }
         \label{mspGem}
   \end{figure}

   \begin{figure}
   \centering
   \includegraphics[width=10cm]{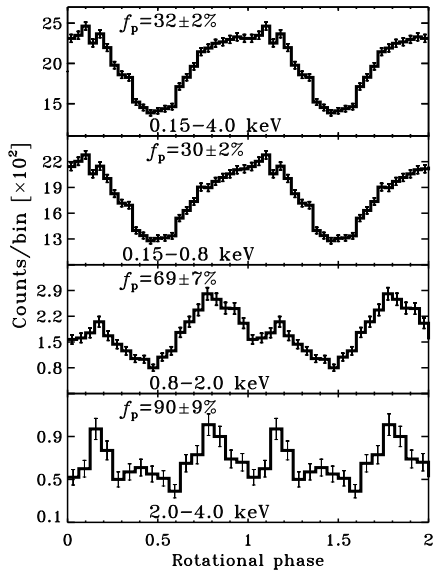}
      \caption{X-ray light curves of Geminga.
              }
         \label{lcGem}
   \end{figure}

A long \xmm\ observation of this famous $\gamma$-ray pulsar in April 2002
revealed a bow-shock nebula 
caused probably by the supersonic motion of the NS
(Caraveo et al.\ 2003). 
The EPIC-pn effective exposure (small-window mode) was 71.4 ks. 
A large number of photons collected 
allows one to accurately measure the spectral 
parameters of the pulsar.
The phase-integrated spectrum nicely fits
a blackbody model, with $T^\infty_{\rm bb}\simeq 0.50$ MK and
$R^\infty_{\rm bb}\simeq 10.6$ km, plus a power-law
component with $\gamma\simeq 1.85$ (Fig.\ 6). 
Similar to B0656, the NS atmosphere models 
yield too large radius-to-distance ratios.
The nonthermal component provides
7\% of the total luminosity in the 0.1--10 keV range\footnote{
The nebula might contribute a small fraction in the estimated nonthermal
flux of the pulsar; a quantitative estimate is expected from a \cha\
observation.},
$L_X^{\rm nonth}\simeq 3\times 10^{30}\,{\rm erg}\,{\rm s}^{-1}
\simeq 1\times 10^{-4}\dot{E}$.
The composite X-ray spectrum extrapolated into the optical domain
(Fig.\ 7) exceeds the observed spectral fluxes by a factor of 50--100, 
whereas the extrapolated thermal component is 
close to the Rayleigh-Jeans spectrum observed in UV (Kargaltsev et al.\ 2004).

Similar to the case of B0656, the X-ray pulse profile 
shows a complex energy dependence 
(Fig.\ 8), with $f_p$ growing up to almost 100\% at
$E>2$ keV, where the emission is purely nonthermal. 
The shape of the light curve extracted in the 2--4 keV range 
is similar to that observed 
in gamma-rays, at $E>100$ MeV (Mayer-Hasselwander et al.\ 1994).
The two narrow peaks of the nonthermal emission are superimposed
on the broad peak at lower energies.
Phase-resolved spectroscopy shows that the photon index
varies with phase from 
$\gamma = 1.7$ at $\phi\simeq 0.4$ (near the minimum of the composite
light curve) to $\gamma = 2.0$ at the phase $\phi\simeq 0.7$ of the 
broader peak in the 2--4\,keV range.

\section{Discussion}
The main contribution to the X-ray fluxes of the three
pulsars is provided by thermal emission 
from the their surfaces. The younger age of J0538 naturally
explains its higher surface temperature, that is
consistent with the standard NS cooling scenario. 
As discussed in ZP02, at such high temperatures ($\gapr 1$\,MK)
and relatively low magnetic fields, the NS surface is expected
to be covered with a gaseous atmosphere,
strongly ionized if comprised of hydrogen. 
Contrary to J0538, the older
B0656 and Geminga are significantly colder and have stronger
magnetic fileds. 
At such conditions, hydrogen atmospheres become denser and less ionized,
and the currently available atmosphere models become less
reliable. On the other hand, diffusive nuclear burning can reduce the
hydrogen abundance in envelopes of cooling NSs (Chang \& Bildsten 2003).
In this case, the 
cooling hydrogen-depleted envelopes may undergo a phase
transition, condensing into a liquid or solid surface.
This may explain why the fits with hydrogen atmosphere
models do not provide reasonable parameters for B0656 and Geminga, 
contrary to the younger and hotter J0538.

The observed temporal behavior of X-ray fluxes from these
three pulsars indicates that their thermal radiation is 
locally anisotropic, in obvious contradiction
with the simplistic blackbody interpretation of the phase-integrated
spectra of B0656 and Geminga. 
Moreover, the asymmetry of the soft X-ray pulses (particularly
for J0538 and B0656) hints that the surface distributions of
temperature and magnetic field are not azimuthally symmetric,
suggesting a strong multipolar component of the magnetic field 
or a decentered magnetic dipole.
Whatever is the true nature of the complicated light curves,
the NS parameters inferred from  the blackbody
spectral fits should be taken with caution. 
In particular, the comparison of the blackbody temperatures
with the predictions of NS cooling theories may not be reliable,
and the blackbody radii may be quite different from the actual 
NS radii.

The X-ray spectra of nonthermal emission from B0656 and Geminga
are close to power laws, with
a phase-dependent photon index in the latter case.
While the power-law slopes and intensities 
(but not the light curves --- Kern et al.\ 2003)
in the optical and X-ray bands are 
in agreement with each other for B0656, they strongly disagree for Geminga. 
It might be explained by the presence of two different populations of
relativistic electrons in the pulsar's magnetosphere.
The younger and more energetic J0538
shows no nonthermal X-ray radiation,
in contradiction with a general expectation that
younger pulsars with larger $\dot{E}$ are stronger nonthermal
emitters. A simplest explanation of this contradiction 
is that the X-ray pulsar beam cannot be seen from the Earth.
Alternatively, J0538 could be one of several pulsars 
with underluminous magnetospheric X-ray emission,
--- for instance, the famous Vela pulsar,
whose age and thermal emission are similar to those of J0538,
shows $L_X^{\rm nonth}/\dot{E}\sim 3\times 10^{-6}$ (Pavlov et al.\ 2001),
well below the upper limit of $2\times 10^{-4}$ inferred for J0538
from the \xmm\ data.

Generally, the new observations confirm that middle-aged pulsars
share a number of common features (in particular, pulsed thermal radiation),
but their X-ray and, especially, multiwavelength emission properties
are rather diverse. Thanks to the much improved quality of X-ray data,
we are now quite certain that we do see the NS surface layers in soft X-rays,
with temperatures $\lapr 1$ MK, generally decreasing with NS age,
and we do see strongly pulsed magnetospheric radiation at higher X-ray energies.
On the other hand, we still do not completely understand 
the mechanisms of thermal
emission (hence, cannot accurately measure the NS temperatures and radii,
despite very small statistical errors), and we lack quantitative models
for magnetospheric emission (hence, cannot infer the complex physics of
NS magnetospheres from the 
measured phenomenological parameters).
However, it is not surprising that we have not yet been able to solve all
the puzzles from observations of a tiny fraction of these exotic
(and diverse!) objects.
The outstanding capabilities of the currently operating observatories,
in both X-rays and optical,
allow observations of a much larger sample of NSs, including thermally emitting
midlle-aged pulsars, and we 
expect that our understanding
of these elusive objects will be much better in a few years, provided enough
observational time is allocated to NS studies.

\begin{acknowledgements}
Work of GGP was partly supported by NASA grant NAG5-10865.
\end{acknowledgements}

\bibliographystyle{aa}

\begin{thebibliography}{}

\bibitem[{Becker \& Aschenbach 2002}]{ba}
Becker, W. \& Aschenbach, B. 2002,
in Proc. of the 270-th Heraeus Seminar on
Neutron Stars, Pulsars and Supernova Remnants, eds. W. Becker, H. Lesch,
J. Tr\"umper, MPE Report 278, 64
\bibitem[{Caraveo et al. 2003}]{ca}
Caraveo, P.A., et al. 2003, Science, 301, 1345
\bibitem[{Chang \& Bildsten 2003}]{cb}
Chang, P. \& Bildsten, L. 2003, ApJ, 585, 464
\bibitem[{Kargaltsev et al. 2004}]{ka}
Kargaltsev, O., et al. 2004, in preparation
\bibitem[{Kern et al. 2003}]{ke}
Kern, B., et al. 2003, ApJ, 597, 1049
\bibitem[{Kramer et al. 2003}]{kr}
Kramer, M., et al. 2003, ApJ, 593, L31
\bibitem[{Mayer-Hasselwander et al. 1994}]{mh}
Mayer-Hasselwander, H.A., et al. 1994, A\&A, 421, 276
\bibitem[{McGowan et al. 2003}]{mc}
McGowan, K.E., et al. 2003, ApJ, 591, 380
\bibitem[{Pavlov et al. 2001}]{p}
Pavlov, G.G., et al. 2001, ApJ, 552, L129
\bibitem[{Pavlov et al. 2002}]{pzs}  
Pavlov, G.G., Zavlin, V.E. \& Sanwal, D., 2002,   
in Proc. of the 270-th Heraeus Seminar on 
Neutron Stars, Pulsars and Supernova Remnants, eds. W. Becker, H. Lesch, 
J. Tr\"umper, MPE Report 278, 273 (PZS02; astro-ph/0206024)
\bibitem[{Romani \& Ng 2003}]{rn}
Romani, R.W. \& Ng., C.-Y. 2003, ApJ, 585, L41
\bibitem[{Weisskopf 2002}]{w}        
Weisskopf, M. 2002,
in Proc. of the 270-th Heraeus Seminar on 
Neutron Stars, Pulsars and Supernova Remnants, eds. W. Becker, H. Lesch, 
J. Tr\"umper, MPE Report 278, 58
\bibitem[{Zavlin \& Pavlov 2002}]{zp}
Zavlin, V.E. \& Pavlov, G.G. 2002,
in Proc. of the 270-th Heraeus Seminar on
Neutron Stars, Pulsars and Supernova Remnants, eds. W. Becker, H. Lesch,
J. Tr\"umper, MPE Report 278, 263 (ZP02; astro-ph/0206025)

\end{thebibliography}

\end{document}